\documentstyle[aps,psfig]{revtex}

\begin{document}
\draft
\title{On the $q$-deformation of the NJL model}
\author{V. S. Tim\'{o}teo and \underline{C. L. Lima}\thanks{%
Corresponding author; e-mail address: cllima@if.usp.br, FAX: +(55) (11) 818
6715.}}
\address{Nuclear Theory and Elementary Particle Phenomenology Group\\
{\em Instituto de F\'{\i }sica, Universidade de S\~{a}o Paulo}\\
{\em CP 66318, 05315-970, S\~{a}o Paulo, SP, Brazil}}
\maketitle

\begin{abstract}
Using a $q$-deformed fermionic algebra we perform explicitly a deformation
of the Nambu--Jona-Lasinio (NJL) Hamiltonian. In the Bogoliubov-Valatin
approach we obtain the deformed version of the functional for the total
energy, which is minimized to obtain the corresponding gap equation. The
breaking of chiral symmetry and its restoration in the limit $q \rightarrow
0 $ are then discussed.
\end{abstract}

\date{\today }
\pacs{PACS numbers: 11.30.Rd, 03.65.Fd, 12.40.-y \\
Keywords: Deformed Algebras, Hadronic Physics, Effective Models}

\vspace*{0.5cm}

One of the most beautiful aspects in physics is the appearance of concepts
which are universal in the sense that they are common to many different
branches in physics. A very important phenomenon in this context is the 
dynamical symmetry breaking, which appears in areas as different as
statistical mechanics or nuclear and particle physics.
In the last few years, results obtained in the treatment of many body 
systems when the underlying algebra is deformed suggests that the 
appearance of symmetry breaking in this new framework might be an 
universal aspect as well. 

The chiral symmetry breaking in Quantum ChromoDynamics (QCD) due to the
appearance of the quark condensates, and its restoration at high temperature
is of fundamental importance in medium and high energy physics. These
mechanisms are responsible for the mass generation of the more fundamental
constituents of matter.

On the other hand, the $q$-deformed algebras have been an alternative and
elegant way to investigate the symmetry breaking process in different areas
of physics \cite{gal,VST}. It has been shown that the $q$-deformation of the
fermionic algebra produces changes in the creation and annihilation operators 
\cite{ubri}. Frequently, physical quantities depend on the action of these
operators in some physical state and therefore they must be sensitive to
changes in the operators' definition.

The aim of this work is to apply this prescription to hadronic systems 
by performing directly the $q$-deformation in the Hamiltonian that 
describes a particular system. 
For this purpose we have chosen the NJL model \cite{NJL}, which has
become a popular effective model for QCD, due to its simplicity and at the
same time the richness in describing some features which are very difficult
to reproduce directly from the more fundamental quantum chromodynamics. For
instance, the dynamical breaking of chiral symmetry and its restoration at
large temperature and density are very well described using such an
effective model \cite{ASYA,VW}.

In a recent work \cite{VST}, the effect of a $q$-deformation in the NJL gap
equation was studied through a $q$-deformed calculation of the quark 
condensates leading to an enhancement of the dynamical mass. 
This can be understood as a
result of a larger effective coupling between the quarks in the deformed
case. In this case, the deformed gap equation is given by 
\begin{equation}
m=-2G\left\langle \overline{\psi }\psi \right\rangle _{q},
\end{equation}
where $\left\langle \overline{\psi }\psi \right\rangle _{q}$ stands for the $%
q$-deformed calculation of the quark condensates.

In this work we use the same deformation procedure as in the previous paper
but, instead of deforming the gap equation, we apply the $q$-deformation in a
more fundamental way by deforming directly the NJL Hamiltonian, which is the
starting point to obtain the gap equation in the variational
Bogoliubov-Valatin approach \cite{K}. Here it is important to note that this
approach is completely different from simply deforming the condensate in the
gap equation, as will be seen in our results.

The Hamiltonian of the Nambu--Jona-Lasinio model is given by 
\begin{equation}
{\cal {H}}_{NJL}=-i\overline{\psi }\gamma \cdot \nabla \psi -G\left[ \left( 
\overline{\psi }\psi \right) ^{2}+\left( \overline{\psi }i\gamma _{5}{\bf %
\tau }\psi \right) ^{2}\right] ,
\end{equation}
and corresponds to the following Lagrangian 
\begin{equation}
{\cal {L}}_{NJL}=\overline{\psi }i\gamma ^{\mu }\partial _{\mu }\psi
+G\left[ \left( \overline{\psi }\psi \right) ^{2}+\left( \overline{\psi }%
i\gamma _{5}{\bf \tau }\psi \right) ^{2}\right] .
\end{equation}
This Lagrangian is constructed from contact interactions in such way that it
has the main symmetries of QCD. As it is well known, one of the most important
features of the QCD Lagrangian is that it has chiral symmetry, which is the
most important symmetry concerning the dynamics of the lightest hadrons.

The $q$-deformed fermionic algebra that we shall use is based on the work of
Ubriaco \cite{ubri}, where the thermodynamic properties of a many fermion
system were studied.  An extension of this procedure was used in the
construction of a $q$-covariant form of the BCS approximation \cite{trip},
and further applied to the NJL gap equation \cite{VST}. As a consequence, this
deformation procedure only modifies negative helicity quarks (anti-quarks)
operators. 

Making use of the $q$-deformed creation and annihilation operators 
we write the modified quark fields as 
\begin{equation}
\psi _{q}(x,0)=\sum_{s}\int \frac{d^{3}p}{\left( 2\pi \right) ^{3}}\left[ B( 
{\bf p},s)u({\bf p},s)e^{i{\bf p\cdot x}}+D^{\dagger }({\bf p},s)v({\bf p}
,s)e^{-i{\bf p\cdot x}}\right] .  \label{bd}
\end{equation}
The $q$-deformed quark and anti-quark creation and annihilation operators 
$B$, $B^{\dagger }$, $D$, and $D^{\dagger }$, are expressed in terms of the
non-deformed ones
\begin{eqnarray}
B_{-} &=& b_{-}\left(1+Qb_{+}^{\dagger }b_{+}\right) \;\;\; , \;\;\;
B_{-}^{\dagger}=b_{-}^{\dagger }\left(1+Qb_{+}^{\dagger }b_{+}\right), 
\label{bm} \\
D_{-} &=& d_{-}\left(1+Qd_{+}^{\dagger }d_{+}\right) \;\;\; , \;\;\;
D_{-}^{\dagger}=d_{-}^{\dagger }\left(1+Qd_{+}^{\dagger }d_{+}\right),
\label{dm}
\end{eqnarray}
where $Q=q^{-1}-1$. 
The positive helicity operators are not modified \cite{VST}.

The variational approach to obtain the gap equation consists on the
following procedure: a) to define a variational vacuum, b) to calculate the
vacuum expectation value of the Hamiltonian, obtaining the functional for
the total energy, and c) to minimize the functional, obtaining the
variational parameters and the gap equation.

Following the above steps, we now define our variational BCS-like vacuum 
\begin{equation}
\left| NJL\right\rangle =\prod_{{\bf p},s=\pm 1}\left[ \cos \theta (p)+s\sin
\theta (p)b^{\dagger }({\bf p},s)d^{\dagger }(-{\bf p},s)\right] \left|
0\right\rangle ,  \label{bcs}
\end{equation}
which, for a given momentum ${\bf p}$, is expanded as 
\begin{eqnarray}
\left| NJL\right\rangle &=&\cos ^{2}\theta (p)\left| 0\right\rangle 
\nonumber \\
&&+\sin \theta (p)\cos \theta (p)b^{\dagger }({\bf p},+)d^{\dagger }(-{\bf p}
,+)\left| 0\right\rangle  \nonumber \\
&&-\sin \theta (p)\cos \theta (p)b^{\dagger }({\bf p},-)d^{\dagger }(-{\bf p}
,-)\left| 0\right\rangle  \nonumber \\
&&-\sin ^{2}\theta (p)b^{\dagger }({\bf p},-)d^{\dagger }(-{\bf p}
,-)b^{\dagger }({\bf p},+)d^{\dagger }(-{\bf p},+)\left| 0\right\rangle .
\end{eqnarray}
Here it is important to note that the deformed version of this vacuum
differs from the non-deformed one only by a phase and, therefore, the
effects of the deformation comes solely from the deformed component of the
Hamiltonian.

The deformed functional for the total energy will be obtained from the
vacuum expectation value of the $q$-deformed NJL Hamiltonian 
\begin{equation}
{\cal W}^{q}\left[ \theta (p)\right] =\left\langle NJL\left| {\cal H}%
_{NJL}^{q}\right| NJL\right\rangle ,
\end{equation}
where 
\begin{equation}
{\cal H}_{NJL}^{q}=-i\overline{\psi }_{q}\gamma \cdot \nabla \psi
_{q}-G\left( \overline{\psi }_{q}\psi _{q}\right) ^{2},
\end{equation}
and $\psi _{q}$ is given by Eq. (\ref{bd}). Due to the additive structure of
the $q$-deformation in Eq. (\ref{bm}) and Eq. (\ref{dm}), the deformed
Hamiltonian can be written as 
\begin{equation}
{\cal H}_{NJL}^{q}={\cal H}_{NJL}+H(Q),  \label{HQ}
\end{equation}
and {\tt the }functional reads 
\begin{equation}
{\cal W}^{q}\left[ \theta (p)\right] ={\cal W}\left[ \theta (p)\right]
+W\left[ Q,\theta (p)\right] .  \label{WQ}
\end{equation}
The last terms of Eqs. (\ref{HQ}) and (\ref{WQ}), namely $H(Q)$ and $W\left[
Q,\theta (p)\right] $, stand for the new terms of first order in $Q$
generated when the algebra is deformed, and therefore, they must vanish for $%
q=1\left( Q=0\right) $. Table \ref{Tab} shows the increase in the number of
operators in the NJL Hamiltonian as also in its matrix elements, due to the
deformation of the fermionic algebra. The hard task is then to find the
non-vanishing matrix elements of the $q$-deformed Hamiltonian.

In the non-deformed case, which corresponds to $q=1\left( Q=0\right) $, the
total energy is given by 
\begin{equation}
{\cal W}\left[ \theta (p)\right] =-2N_{c}N_{f}\int \frac{d^{3}p}{\left( 2\pi
\right) ^{3}}p\cos 2\theta (p)-4G\left( N_{c}N_{f}\right) ^{2}\left[ \int 
\frac{d^{3}p}{\left( 2\pi \right) ^{3}}\sin 2\theta (p)\right] ^{2}.
\end{equation}
The minimization of this functional 
\begin{equation}
\frac{\delta {\cal W}\left[ \theta (p)\right] }{\delta \left[ 2\theta
(p)\right] }=0,
\end{equation}
leads to the NJL gap equation 
\begin{equation}
p\tan 2\theta (p)=4GN_{c}N_{f}\int \frac{d^{3}p^\prime}{\left( 4\pi
\right)^{3}} \sin 2\theta (p^\prime),  \label{bvgap}
\end{equation}
which takes its more familiar form 
\begin{equation}
m=4GN_{c}N_{f}\int \frac{d^{3}p}{\left( 2\pi \right) ^{3}} \frac{m}{\sqrt{%
{\bf p}^{2}+m^{2}}},  \label{gap}
\end{equation}
provided the variational angles acquire the following structure 
\begin{equation}
\tan 2\theta (p)=\frac{m}{p},\sin 2\theta (p)=\frac{m}{\sqrt{{\bf p}
^{2}+m^{2}}},\cos 2\theta (p)=\frac{p}{\sqrt{{\bf p}^{2}+m^{2}}}.
\label{ang}
\end{equation}

Calculating the new matrix elements arising from the $q$-deformation of the
NJL Hamiltonian, and adding them up to the non-deformed functional, we
obtain the full $q$-deformed functional for the total energy 
\begin{equation}
{\cal W}^q\left[ \theta (p)\right] =-2N_cN_f\int \frac{d^3p}{\left( 2\pi
\right) ^3}P_q\cos 2\theta (p)-4G^{^{\prime }}\left( N_cN_f\right) ^2\left[
\int \frac{d^3p}{\left( 2\pi \right) ^3}\sin 2\theta (p)\right] ^2.
\label{full}
\end{equation}
As in the non-deformed case, the same minimization procedure yields to 
\begin{equation}
P_q\tan 2\theta (p)=4G^{^{\prime }}N_cN_f\int \frac{d^3p^{\prime }}{\left(
4\pi \right) ^3}\sin 2\theta (p^{\prime }),  \label{qbvgap}
\end{equation}
which becomes 
\begin{equation}
M=4G^{^{\prime }}N_{c}N_{f}\int \frac{d^{3}p}{\left( 2\pi \right) ^{3}} 
\frac{M}{\sqrt{{\bf P}^{2}_q + M^{2}}}.  \label{qgap}
\end{equation}
The variational angles have the same old structure but now are $q$-dependent 
\begin{equation}
\tan 2\theta_{q}(p)=\frac{M}{P_{q}},\sin 2\theta_{q}(p)= \frac{M}{\sqrt{{\bf %
P}_{q}^{2}+M^{2}}}, \cos 2\theta _{q}(p)=\frac{P_{q}}{\sqrt{{\bf P}
_{q}^{2}+M^{2}}},  \label{qang}
\end{equation}
where the new variables appearing in the deformed equations are defined as 
\begin{eqnarray}
P &=& \left( 1 + \frac{Q}{2} \right) p , \\
P_{0} &=& \frac{N_{c}N_{f}}{3\pi^{2}}\frac{Q}{2}\,G\Lambda ^{3} , \\
P_{q} &=& P-P_{0} , \\
G^{^{\prime }} &=& G\left( 1+\frac{Q}{4}\right) .
\end{eqnarray}
It is easy to see that, when $q\rightarrow 1\left( Q\rightarrow 0\right) $,
Eqs. (\ref{qbvgap}), (\ref{qgap}), and (\ref{qang}) reduce to their
non-deformed versions Eqs. (\ref{bvgap}), (\ref{gap}), and (\ref{ang}),
since $P\rightarrow p $, $P_0\rightarrow 0$, $P_q\rightarrow p$, and $%
G^{^{\prime }}\rightarrow G$.

In analogy with the non-deformed case we can write the gap equation in terms
of the quark condensates as 
\begin{equation}
M = -2 G^{^{\prime }} \left\langle \overline{\Psi} \Psi \right\rangle .
\label{gapqq}
\end{equation}
Comparing the two forms of the gap equation Eqs. (\ref{qgap}) and (\ref
{gapqq}), we find a new deformed condensate given by 
\begin{equation}
\left\langle \overline{\Psi} \Psi \right\rangle = - \frac{N_c}{\pi^2}%
\int_0^\Lambda dp\, p^2 \frac{M}{\sqrt{{\bf P}_q^2 + M^2}}
\end{equation}
for each quark flavour. This condensate is different from the one obtained
in our previous work, where the condensate was explicitly deformed \cite
{VST}. It also has exactly the same form of the non-deformed one, but is
written in terms of the new variables. It is worth to mention that the new
condensate is not obtained by calculating the vacuum expectation value of a
deformed scalar density, it corresponds to the gap equation which arises
from the variational procedure started from the $q$-deformed Hamiltonian.

We can also obtain a new pion decay constant in analogy to the non-deformed
case \footnote{In Ref. \cite{VST} the factor $N_c m^2$ is missing in Eq. (24).}
\begin{equation}
F_{\pi }^{2}=N_{c}M^{2}\int_{0}^{\Lambda }\frac{d^{3}p}{\left( 2\pi \right)
^{3}}\frac{1}{\left( {\bf P_{q}}^{2}+M^{2}\right) ^{3/2}}.
\end{equation}

In the non-deformed case, a coupling constant equal to $G/G_{c}=0.75$ leads
to a pion decay constant $F_{\pi }=88$ MeV. By setting the deformation
parameter to $q=1.2$, the pion decay constant is shifted to $F_{\pi }=92$
MeV which is very close to experimental value $f_{\pi }=93$ MeV. This is in
agreement with the results obtained in the context of chiral perturbation
theory by Gasser and Leutwyler \cite{gasser}, where the calculated pion
decay constant is reduced by $6\%$ becoming $F_{\pi }=88$ MeV when current
quark masses $m_{u,d}$ are set to zero.

The variational angle plays an important role in the chiral symmetry
breaking process. When $2\theta =0$, there is no breaking of chiral symmetry
and therefore no dynamical mass is generated. The generation of the
dynamical mass is associated to a chiral rotation and the presence of a
current quark mass can be associated to a non-vanishing angle, namely to a
permanent chiral rotation. The study of the behavior of the variational
angle provides an interesting way to observe the dynamical chiral symmetry
breaking in the Bogoliubov-Valatin approach to the NJL model.

\begin{figure}[t]
\centerline{\psfig{figure=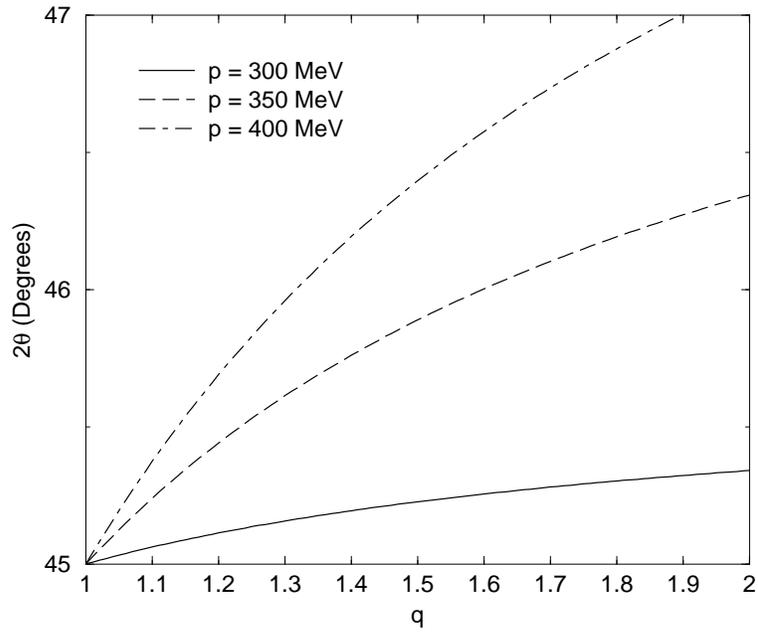,width=10cm}}
\caption{The variational angle $2\theta$ as a function of the deformation
parameter $q$ when the dynamical mass is 300 MeV.}
\label{Txq}
\end{figure}

\begin{figure}[b]
\centerline{\psfig{figure=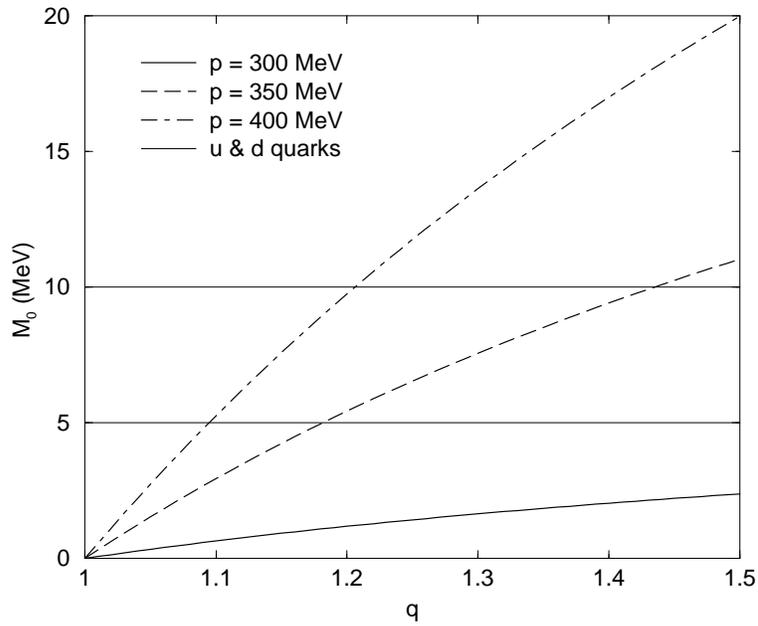,width=10cm}}
\caption{The corresponding generated mass as a function of the deformation
parameter $q$ when the momentum is fixed. When $M_0 = 0$ the dynamical mass
is 300 MeV.}
\label{Mxq}
\end{figure}

By fixing the dynamical mass we can use Eqs. (\ref{qang}) to obtain the $q$
-dependence of the variational angle, which is shown in Fig. (\ref{Txq}) for
different values of the momentum. Then, starting from the fixed value of the
dynamical mass, we use the $q$-dependence of $2\theta$ to obtain the mass
generated when the fermionic algebra is deformed. In Fig. (\ref{Mxq}) we
show the difference between the fixed value of the dynamical mass and the 
$q$-dependent one, compared to typical values for the current $u$ and $d$
masses.

\begin{figure}[t]
\centerline{\psfig{figure=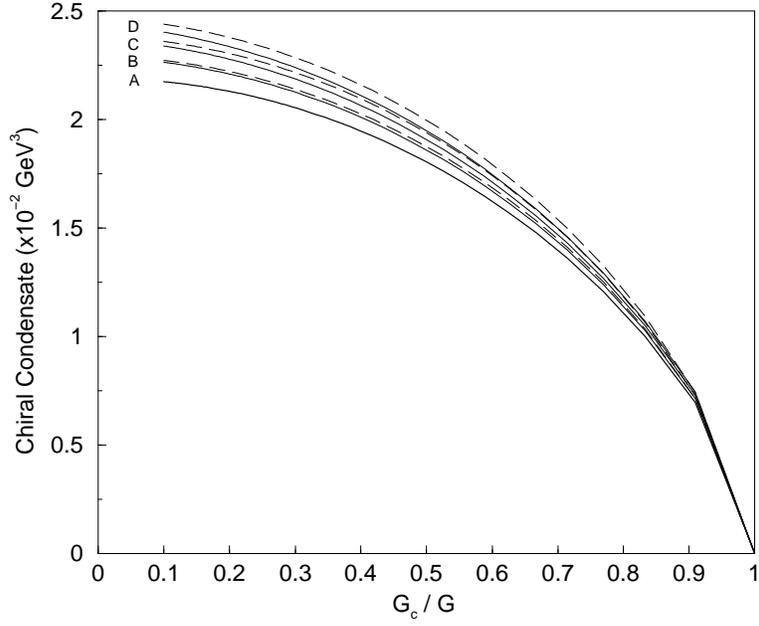,width=10cm}}
\caption{The NJL phase transition. Four sets of curves are presented.
The correspond to: (A) the non-deformed case $q=1$, (B) $q=1.1$, 
(C) $q=1.2$, and (D) $q=1.3$. In each set (B,C,D) the lower curve 
(continuous line) is obtained by deforming directly the gap equation,
and the upper one (dashed line) corresponds to the results obtained from 
the deformation of the NJL Hamiltonian.}
\label{Qfunc}
\end{figure}

\begin{figure}[b] 
\centerline{\psfig{figure=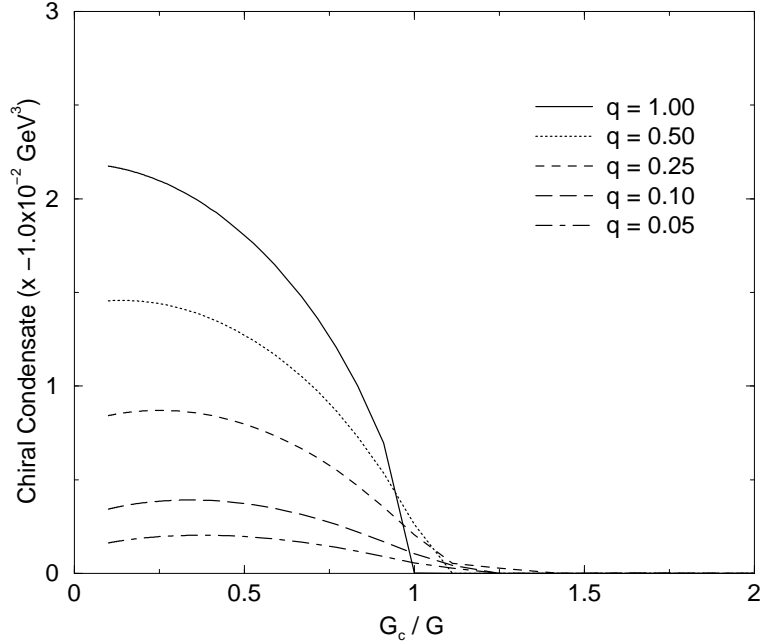,width=10cm}} 
\caption{The condensate for $0 < q < 1$.} 
\label{Qlt1} 
\end{figure}

\begin{figure}[t] 
\centerline{\psfig{figure=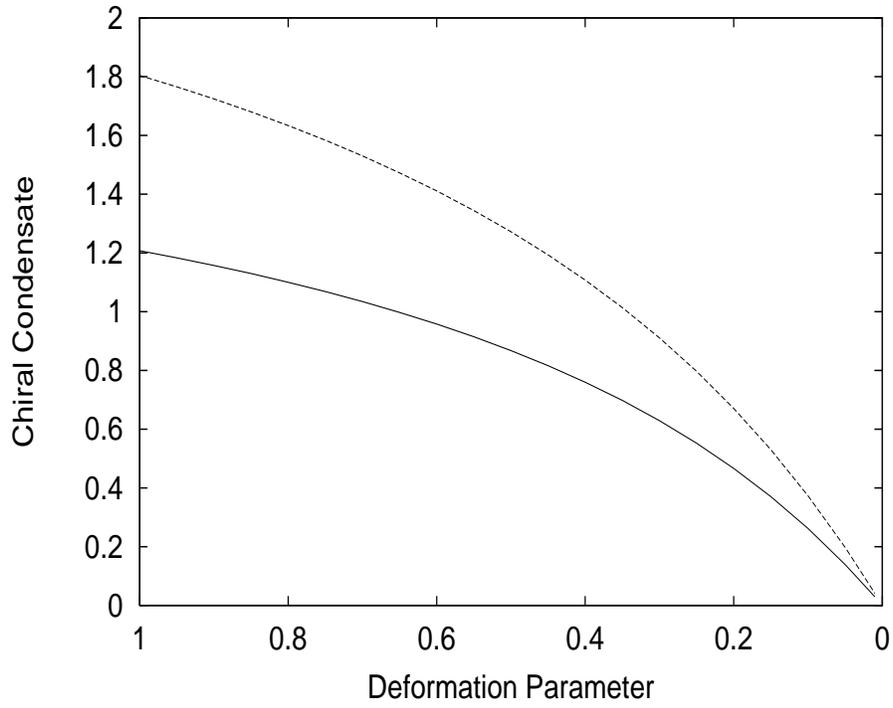,height=10cm,width=12cm}} 
\caption{Chiral symmetry restoration as $q \rightarrow 0$. The solid and
dashed lines correspond to $G=5.94$ and $G=9.14$ respectively.}  
\label{rest}  
\end{figure}

The NJL phase transition can be studied by solving the gap equation Eq. (\ref
{qgap}) and calculating the quark condensate $\left\langle \overline{\Psi }%
\Psi \right\rangle $, which is the chiral order parameter. If we solve this $%
q$-deformed self-consistent equation in terms of the new variables defined
above $\left( P_{q},\,E_{q}\right) $ the phase transition will look exactly
like in the non-deformed case, since Eq. (\ref{qgap}) is identical to usual
gap equation Eq. (\ref{gap}). However, if we solve it in terms of the old
variables $\left( p,\,E\right)$ we can see the effect of the $q$-deformation 
in the phase transition. The curves shown in Fig. (\ref{Qfunc}) for 
different values of $q$ are compared to the previous approach, where the 
$q$-deformation was performed only in the quark condensates \cite{VST}. 
This feature can be understood as follows. We have two separated scenarios: 
the non-deformed and the $q$-deformed one. In both situations the gap 
equation has exactly the same form and yields the same results. 
The effects of the deformation are observed when we express the gap 
equation of the deformed case in terms of the original physical quantities 
of the non-deformed one. For $q>1$ the
value of the quark condensates and the dynamical mass increase with the
deformation and are larger than in the case where only the quark 
condensates were
deformed. For $q\lesssim $ $1$ we have the opposite effect and the value of
the condensate decreases. It is therefore tempting to explore the 
behavior of the condensate for smaller values of $q$, 
even considering that the
truncation at order $Q$ may not be granted. In this case, we can see that
the chiral symmetry is restored in the limit $q\rightarrow 0$, since the
condensate vanishes. The value of the condensate for $q<1$ is shown in Fig. (
\ref{Qlt1}), and in Fig. (\ref{rest}) we can see the chiral symmetry
restoration at small values of the deformation parameter $q$ at fixed value
of the coupling constant.

The chiral symmetry restoration here is different from the obtained at
finite temperature \cite{ASYA,VW}. It seems to be important an investigation
of the effect of both temperature and $q$-deformation in the chiral symmetry
restoration process. This study is in progress and will be left for a future
publication \cite{next}.

So far we have performed the $q$-deformation of the NJL Hamiltonian and
used with the variational Bogoliubov-Valatin approach to obtain the $q$
-deformed functional that leads to a new gap equation. As far as effects of
the deformation are concerned, our main conclusions can be summarized as
follows.

In this approach the variational angles become $q$-dependent, meaning that
the dynamical mass generation is affected by the $q$-deformation. The effect
of the deformation is to enhance the condensate and the dynamical mass for $%
q > 1$, and to restore chiral symmetry when $q \rightarrow 0$. Here the
effect is stronger than in the case where the deformation is performed
directly into the gap equation. In terms of the new $q$-deformed variables
the functional for the total energy, the gap equation, the variational
angles, and the quark condensates have the same form of the non-deformed
case, which is a consequence of the quantum group invariance of the NJL
Lagrangian.

{\bf Acknowledgments}

The authors are grateful to U. Mei\ss ner and M. R. Robilotta for the
suggestion which motivated our study on the $f_\pi$ behaviour.
C. L. L. is grateful to D. Galetti and B. M. Pimentel for very helpful
discussions, and V. S. T. would like to acknowledge FAPESP for financial
support. This work was supported by FAPESP Grant Nos. 98/6590-2 and 98/2249-4.


\begin{table}[tbp]
\centering
\begin{tabular}{ccccc}
${\cal H}_{NJL}$ & {\bf Without} & {\bf With} & {\bf New} & {\bf Order} $%
{\bf Q}$ \\ 
$-i\overline{\psi }\gamma \cdot \nabla \psi $ & {16} & {36} & {20} & {16} \\ 
$-G\overline{\psi }\psi \overline{\psi }\psi $ & {256} & {1296} & {1040} & {%
520} \\ 
$\left\langle NJL\left| -i\overline{\psi }\gamma \cdot \nabla \psi \right|
NJL\right\rangle $ & {256} & {576} & {320} & {256} \\ 
$\left\langle NJL\left| -G\overline{\psi }\psi \overline{\psi }\psi \right|
NJL\right\rangle $ & {4096} & {20736} & {16640} & {8320}
\end{tabular}
\caption{Number of terms of the Hamiltonian and its matrix elements {\bf %
without} deformation, {\bf with} deformation, the {\bf new} terms, and the
terms of {\bf first order in} $Q$. }
\label{Tab}
\end{table}


\end{document}